\documentclass[conference,10pt,top=0.7in]{IEEEtran}

\usepackage{graphicx,color}
\usepackage{amsmath, amsthm, amsfonts, amssymb, amsbsy,nccmath}
\usepackage{mathtools}
\usepackage{algorithm}
\usepackage{algcompatible}
\usepackage{enumerate}
\usepackage{lipsum}
\newtheorem{lemma}{Lemma}
\usepackage{textgreek}
\usepackage{textcomp}
\usepackage{stfloats}

\usepackage{algpseudocode}

\usepackage[sort,compress]{cite}
\usepackage{epsfig}
\usepackage{epstopdf}
\usepackage{mathtools}
\usepackage{dsfont}
\usepackage{epstopdf}

\usepackage{sidecap, caption}
\usepackage{subcaption}
\theoremstyle{definition}
\newtheorem{definition}{Definition}

\theoremstyle{assumption}
\newtheorem{assumption}{Assumption}[section]

\theoremstyle{proposition}

\theoremstyle{corollary}

\usepackage[inline]{enumitem}   
\makeatletter
\newcommand{\inlineitem}[1][]{%
\ifnum\enit@type=\tw@
    {\descriptionlabel{#1}}
  \hspace{\labelsep}%
\else
  \ifnum\enit@type=\z@
       \refstepcounter{\@listctr}\fi
    \quad\@itemlabel\hspace{\labelsep}%
\fi}
\makeatother
\usepackage{comment}

\newtheorem{theorem}{Theorem}

\newcommand{\beq}{\begin{equation}}
\newcommand{\eeq}{\end{equation}}

\DeclareMathOperator*{\argmax}{arg\,max}
\DeclareMathOperator*{\argmin}{arg\,min}

\newcommand{\dsum}{\displaystyle\sum}

\def\adots{\mathinner{\mskip0mu\raise0pt\vbox{\kern7pt\hbox{.}}\mskip3mu
          \raise4pt\hbox{.}\mskip3mu\raise8pt\hbox{.}\mskip0mu}}

\newcommand{\bmW}{{\boldsymbol W}}

\usepackage{bm}

\newcommand{\bmx}{{\bm x}}

\newcommand{\bmw}{{\bm w}}

\newcommand{\bmG}{{\bm G}}

\newcommand{\mS}{\mathcal{S}}
\newcommand{\mF}{\mathcal{F}}

\newcommand{\mX}{\mathcal{X}}
\newcommand{\mH}{\mathcal{H}}

\newcommand{\mJ}{\mathcal{J}}

\newcommand{\mbE}{\mathbb{E}}

\newcommand{\mO}{\mathcal{O}}
\newcommand{\mA}{\mathcal{A}}
\newcommand{\mQ}{\mathcal{Q}}
\newcommand{\mU}{\mathcal{U}}
\newcommand{\mW}{\mathcal{W}}
\newcommand{\mR}{\mathcal{R}}
\newcommand{\mE}{\mathcal{E}}

\newcommand{\mN}{\mathcal{N}}

\newcommand{\mT}{\mathcal{T}}
\newcommand{\mM}{\mathcal{M}}

\newcommand{\bpi}{\boldsymbol{\pi}}
\newcommand{\bpih}{\widehat{\bpi}}

\newcommand{\bma}{{\bm a}}

\newcommand{\bTheta}{\boldsymbol{\Theta}}

\newcommand{\btheta}{\boldsymbol{\theta}}

\newcommand{\bmah}{\widehat{\bma}}

\makeatletter
\newcommand\fs@spaceruled{\def\@fs@cfont{\bfseries}\let\@fs@capt\floatc@ruled
  \def\@fs@pre{\vspace{0.5\baselineskip}\hrule height.8pt depth0pt \kern2pt}%
  \def\@fs@post{\kern1pt\hrule\relax}%
  \def\@fs@mid{\kern2pt\hrule\kern2pt}%
  \let\@fs@iftopcapt\iftrue}
\makeatother

\newcommand{\bit}{\begin{itemize}}
\newcommand{\eit}{\end{itemize}}

\newcommand{\mK}{\mathcal{K}}
\newcommand{\mL}{\mathcal{L}}

\newcommand{\mG}{\mathcal{G}}
\newcommand{\mC}{\mathcal{C}}
\newcommand{\mCh}{\widehat{\mC}}

\newcommand{\bms}{{\boldsymbol s}}

\DeclarePairedDelimiter\abs{\lvert}{\rvert}%

\usepackage{lipsum}

\makeatletter

\newcommand\longleftrightarrowfill@{%
  \arrowfill@\leftarrow\relbar\rightarrow}
\makeatother

\restylefloat{algorithm}

\IEEEoverridecommandlockouts
\begin{document}

\title{\vspace{4mm}Hypergame Theory for Decentralized Resource Allocation in Multi-user Semantic Communications \vspace{-6mm}}
\vspace{-6mm}
\author{\fontsize{1}{1}\selectfont
\IEEEauthorblockN{\fontsize{11}{11}\selectfont
Christo Kurisummoottil Thomas and Walid Saad\thanks{This research was supported by the Office of Naval Research (ONR) under MURI grant N00014-19-1-2621.}
\vspace{0mm}}
Wireless@VT, Bradley Department of Electrical and Computer Engineering, \\ Virginia Tech, Arlington, VA, USA.\\ \fontsize{8}{8}Emails: \{christokt,walids\}@vt.edu\vspace{-7mm}
}
\maketitle

\vspace{-1mm}
\begin{abstract}\vspace{-0mm}
Semantic communications (SC) is an emerging communication paradigm in which wireless devices can send only relevant information from a source of data while relying on computing resources to regenerate missing data points. However, the design of a multi-user SC system becomes more challenging because of the computing and communication overhead required for coordination. Existing solutions for learning the semantic language and performing resource allocation often fail to capture the computing and communication tradeoffs involved in multi-user SC. To address this gap, a novel framework for decentralized computing and communication resource allocation in multi-user SC systems is proposed. 
The challenge of efficiently allocating communication and computing resources (for reasoning) in a decentralized manner to maximize the quality of task experience for the end users is addressed through the application of Stackelberg hypergame theory. Leveraging the concept of second-level hypergames, novel analytical formulations are developed to model misperceptions of the users about each other's communication and control strategies. Further, equilibrium analysis of the learned resource allocation protocols examines the convergence of the computing and communication strategies to a local Stackelberg equilibria, considering misperceptions. Simulation results show that the proposed Stackelberg hypergame results in efficient usage of communication and computing resources while maintaining a high quality of experience for the users compared to state-of-the-art that does not account for the misperceptions.
\end{abstract}\vspace{-1mm}

\vspace{-2mm}
\section{Introduction}
\label{section_intro}
\vspace{-2mm}

Semantic communications (SC) is a promising approach to enhance transmission efficiency in future wireless networks, such as 6G, by harnessing the reasoning capabilities of end users and exploiting the structure of wireless data \cite{ChaccourArxiv2022}.  The benefits of SC extend across multiple layers of the open systems interconnection (OSI) networking model. These include semantic extraction by both the transmitter and receiver, leveraging extracted semantics to guide computing and communication resource allocation, and managing quality of service. Transmission efficiency in SC is achieved by transmitting the learnable structure present in the data. At the receiver side, generalizable artificial intelligence (AI) models \cite{SaadProceedings2024} can be used to perform reasoning the data points that are missed due to bad channel quality or any network disruptions. However, in multi-user systems, the user quality of experience may be impacted due to delays in computing and communication. Delays may occur while waiting for reliable communication links to be established or for access to shared computing resources.

\vspace{-2mm}\subsection{Related Works}\vspace{-1mm}

Majority of the prior works in SC \cite{LiuTCCN2023,YanGC2022,ZhaoArxiv2024} are limited to optimizing traditional physical layer functions such as channel assignment, power allocation, and transmit symbols. Moreover, the work in \cite{LiuTCCN2023} presented an approach for dynamically optimizing  the data features to be communicated based on their relevance to the end-user. 
However, the resource allocation schemes in \cite{LiuTCCN2023} and \cite{YanGC2022} fail to account for the possibility that end-users may possess reasoning capabilities.  Reasoning AI models \cite{SaadProceedings2024} enable communication nodes to infer missing variables and improve the prediction of future events. Integrating reasoning capabilities into the resource allocation problem could enhance resource utilization while meeting the demanding delay and throughput expected of future wireless systems. Although the authors in \cite{ZhaoArxiv2024} explored the use of symbolic AI techniques to perform multi-user resource allocation, their contribution is again limited to traditional tasks such as uplink or downlink channel assignment. 
Another drawback of existing multi-user resource allocation schemes \cite{YanGC2022,LiuTCCN2023,ZhaoArxiv2024} is their assumption that the semantic relevance of features in the data is known to the transmitters. Additionally, these schemes often require joint training of deep learning modules across the transmit and receive nodes which can lead to significant communication overhead. Moreover, extensive retraining efforts are required when network or channel environments change, which can be cumbersome. 
In contrast to the state-of-the-art \cite{YanGC2022,LiuTCCN2023,ZhaoArxiv2024}, semantic-aware resource allocation must guarantee high semantic reliability by efficiently allocating communication and computing resources in a decentralized manner. 
\vspace{-2mm}\subsection{Contributions}\vspace{-1mm}
The main contribution of this paper is a rigorous framework for decentralized computing and communication resource allocation in a multi-user wireless SC system using hypergame theory. In particular, hypergames allow to analyze the \emph{hyper Stackelberg equilibrium (HSE)} \cite{KovachGT2015} when the players involved in the game may have
incorrect beliefs about the other players’ strategies, or preferences.
We propose a Stackelberg hypergame model to optimize computing and communication resource allocation policies among a set of transmitters (TX, leaders) and receivers (RX, followers) in a decentralized manner. Our computing model accounts for the RXs' ability to utilize either local computing resources or shared cloud computing (CC) resources for reasoning to infer missing information caused by degraded channel quality.
We propose a swap learning method \cite{GharesifardCDC2010} to update each user's perceptions of other users' communication or computing strategies and the RX's semantic relevance factors.
We conduct an equilibrium and stability analysis of the proposed hypergame model for the case of a two-user system with a single leader and a single follower. Furthermore, we prove the existence of a local HSE for the two-user system when misperception occurs at both the TX and RX, and the swap learning-based perception updates converge. 
    The equilibrium strategies reveal that integrating \emph{semantic relevance} and RX \emph{reasoning capabilities} into TX strategies leads to a decrease in the number of bits transmitted across the network. Furthermore, receivers can effectively allocate computing resources to minimize the \emph{reasoning success probability} by accurately perceiving other RX strategies. Simulation results show nearly $45\%$ reduction in the number of physical bits communicated compared to classical systems that do not incorporate reasoning capabilities at the RXs. Moreover, the proposed scheme reduces the gap to complete information games by approximately $18\%$ in terms of the quality-of-task-experience (QoTE) at the RXs.

\vspace{-1mm}\section{System Model}
\vspace{-1mm}

Consider a multi-user communication system involving $K$ users. We consider that $K/2$ of the users are transmitters, with each TX $k$ equipped with sensors that allow them to observe a part of a surrounding physical environment $\bmx_k \subset \mR^N$. Each TX  intends to communicate the observations to the other $K/2$ users that are RXs. The entire wireless environment can be captured using a global set of observations $\mX = \{\bmx_1,\cdots,\bmx_{\frac{K}{2}}\}$. The set of TXs is defined as $\mK$ and the RX set is $\mJ$.  The receivers perform an RX-specific task $T_j \in \mathcal{T}_j$. An example of such a system is a group of autonomous robots in a smart factory, where each robot is assigned a specific task and receives rewards for completing those tasks. In this scenario, the robots in the smart factory represent the RXs. The diverse sensing elements located throughout the smart factory serve as the TXs. Additionally, each RX might have its own observations through a head-mounted mixed reality (MR) display. In this scenario, to efficiently execute task $T_j$,  RX $j$ requires the global observation $\mX$. 
 A naive approach here is to let every TX communicate simultaneously over a shared broadcast channel. 
 However, the users operate in a communication and computing constrained environment. Communication constraints mean that due to poor channel conditions caused by multi-user interference or the absence of line-of-sight links  \cite{ChaccourITJ2022}, some of the communicated information may not be reliably decoded. To address such extreme scenarios, we assume that the RXs are equipped with \emph{computing capabilities enabling them to reason about missing information communicated}.  To facilitate reasoning, we assume that the RXs acquire background causal knowledge about their environment through emergent language training, as we showed in our work in \cite{ChristoTWCArxiv2022}. Given the reasoning capability of the RXs, TXs have the flexibility to communicate only the relevant information and cannot be reasoned, thus enhancing transmission efficiency. However, due to resource constraints, the system must balance between communication and computing, motivating the use of SC. In the considered SC system, each RX aims to: (a) successfully interpret relevant information within a set timeframe and (b) maximize the \emph{(QoTE)}.  However, given the constraints on the computing and communication resources, the utility function (i.e., QoTE-based) of any TX or RX on the communication and computing strategies of other TXs and RXs. Next, we define the semantics reasoning model.

\vspace{-2mm}\subsection{Semantics Reasoning Model}\vspace{-1mm}

The local observation $\bmx_k$ at any TX $k$ is generated using a function $O_k: \{\mC_k,\mE_k\} \rightarrow \mX_k$, where $\mX_k$ is the local observation space, $\mC_k$ is the set of features relevant to any RX task, and $\mE_k$ is the set of features that are irrelevant to any of the RX tasks. The TX extracts the relevant features $\mC_k$ from the observed data $\bmx_k$, defined as the \emph{semantic concepts}, represented by a set $\mC_k = \{c_{k1},c_{k2},\cdots, c_{kD}\}$. 
Unlike statistics-based compression, semantic concept extraction here emphasizes the causal relationships among concepts and how these relationships enable RXs to draw logical conclusions relevant to task execution. A causal graph $G_k$ describing these relationships among $\mC_k$ can be learned using techniques like generative flow networks, as done in \cite{ChristoTWCArxiv2022}. 
Here, we focus on how the knowledge of the semantic relevance of concepts to RX tasks helps efficiently allocate computing and communication resources. 
We define the global set of semantic concepts as $\mC = \cup_{k\in \mK} \mC_k$, with the corresponding global observation $\mO :\{\mC,\mE\} \rightarrow \mX$.  However, for reliably executing a task at RX $j$, this RX may only need to know a subset $\mC_{k}^{(j)} \subseteq \mC_k, \forall k\in \mK$, with $\abs{\mC_{k}^{(j)}} \leq D$.  Whenever the communication link from TX $k$ to RX $j$ is not decodable, RX $j$ can perform reasoning to deduce the semantic concepts TX $k$ intended to communicate. Reasoning here is accomplished using interventions performed on the causal graph \cite{PearlBasic2018}. For any $c_{kr}$ that should be reasoned at RX $j$, the corresponding intervention can be formulated as computing a posterior belief about the missing concepts given an incomplete $G_j$ formed by the reliably decoded concepts:
\vspace{-1mm}\beq
\vspace{-1mm}\begin{aligned}
c_{kr}^{j\,\ast} = \argmax\limits_{c_{kr}^0} p(\{\widehat{\mC}_k^{(j)}\backslash c_{kr}^{(j)}\}\mid \textrm{do}(c_{kr}) = c_{kr}^0).
\end{aligned}
\label{eq_interventions}
\vspace{-0mm}\eeq
\eqref{eq_interventions} means inferring the semantic concepts from TX $k$ that best explain the remaining concepts, $\{\widehat{\mC}_k^{(j)}\backslash c_{kr}^{(j)}\}$. To compute \eqref{eq_interventions}, the RX can employ causal Bayesian optimization \cite{AgliettiPMLR2020}. Further, we represent each task $T_j$ by a tuple,  $\left(\mW_j,\mF_j\right)$, where $\mW_j$ is a set that contains the relevance factor (defined as the possible weight vectors $\bmw_k^{(j)} = \left[w_{kr}^{(j)},\forall r\right]\in [0,1]^D$) of each semantic concept $c_{kr}$ extracted at TX $k$. As studied in \cite{ThomasGC2022}, the task description at any RX $j$ can be quantified using a set of logical formulas (specifically symbolic functions) that it must evaluate. Each such symbolic function can be defined using a set of semantic concepts and the connectives $\{\lor,\land,\neg,\!\implies\!, \!\iff\!,\!\Leftrightarrow
\}$. $w_{kr}^{(j)}$ can be precisely defined as the fraction of logical formulas that constitute the semantic concept $c_{kr}$. The semantic relevance factor $w_{kr}^{(j)}$ determines the subset of semantic concepts $\mC_{k}^{(j)}$ that must be communicated over the link from TX $k$ to RX $j$. However, $w_{kr}^{(j)}$ is unknown to the TX and must be learned. $\mF_j$ is a set that contains the number of computations (in cycles) $F_{kr}^{(j)}$ required per reasoning a semantic concept. $F_{kr}^{(j)}$ can be different across RXs for the same concept $c_{kr}$ due to the different levels of RX cognition and computing capabilities.  The accuracy of the causal graph available at each RX, that is obtained through the causal discovery models \cite{ChristoTWCArxiv2022}, defines the RX cognition here. Based on this causal knowledge, the number of interventions to be performed for a specific semantic concept will vary, and this maps into a varying computation requirements $F_{kr}^{(j)}$ per concept. 

\vspace{-2mm}\subsection{Computing Model}
\vspace{-2mm}

To perform the reasoning computations \eqref{eq_interventions}, RX $j$ can use either the limited, locally available computing resources or the more significant resources at a CC server. The local computing resources are limited to performing $F_j^{\mathrm{max}}$ cycles per sec (cycles/s) $,\forall j$. The CC server resources can perform $F_0^{\mathrm{max}} \gg F_j^{\textrm{max}}$ cycles/s, but it is shared between all RXs. We define $d_j$ as the number of computing cycles/s reserved by the CC server for any RX $j$. Another disadvantage of using a CC server for computing is that the total delay involved in obtaining the results may be larger due to the extra communication involved between RX and CC server. The delay incurred by performing reasoning for any semantic concept $c_{kr}$ using a CC server or a local server is $\frac{a_{l,kr}}{R_j} + \frac{\sum\limits_r F_{kr}^{(j)}}{d_{j}}$, where $a_{l,kr}$ is defined as the number of bits needed to represent $c_{kr}$ and $R_j= B\log_2(1+\frac{\abs{h_j}P_j}{N_0B})$ is the rate of the link between RX $j$ and the CC server, 
where $\abs{h_j}$ is the effective channel gain between RX $j$ and the CC server, $P_j$ is the power allocated (considered as fixed), $B$ is the allocated bandwidth which is assumed to be the same for all users, and $N_0$ is the noise power spectral density. The following assumptions are considered for our subseuqent analysis.
\vspace{-2mm}\begin{assumption}
    $h_j$ includes the effect of multi-user beamforming, and the links to the CC server are assumed to operate under zero or negligible inter-user interference (high signal-to-noise-ratio regime). 
\end{assumption}
\vspace{-3mm}\begin{assumption}
\label{assumption_2}
    The outcomes of both computations and transmissions (defined as the vector $\boldsymbol{s} \in \mathcal{S}$), encompassing information regarding the dropped semantic concepts and the number of physical bits communicated from each TX, are accessible to all users.
\vspace{-2mm}\end{assumption}
Assumption~\ref{assumption_2} is practically feasible, as information about dropped packets can be fed back via control channels, incurring minimal communication overhead.

\vspace{-2mm}\section{Hypergame Formulation for Communication and Computing Resource Allocation}
\vspace{-2mm}

Given the set $\mC_k$ and the channel quality captured via a distribution $p(\widehat{c}_{kr}^{(j)}\mid c_{kr})$ ($\widehat{c}_{kr}^{(j)}$ is the decoded concept at RX $j$), the TX's strategy is to encode the semantic concepts using a certain number of physical bits that should be transmitted. We assume that distinct concepts from any TX are transmitted across orthogonal channels, but there could be interference between the concepts from different TXs (for example, between $c_{kr}$ and $c_{ir}$). The resulting interference and fading are captured using $p(\widehat{c}_{kr}^{(j)}\mid c_{kr})$.  To compute an efficient communication strategy that minimizes the physical bits transmitted, each TX must know the semantic relevance of the extracted concepts to each RX tasks. This semantic relevance factor enables TXs to allocate the number of  sufficient bits $a_{l,kr}$ for each relevant concept and to not communicate those which are irrelevant to the RXs. Similarly, each RX requires information on whether other RXs rely on reasoning computations or communicated semantic concepts for optimal decision-making. Due to the intricate inter-dependencies among TX and RX choices, modeling the problem using game theory is a promising approach. We pose our problem as a multi-leader, multi-follower Stackelberg game, in which the leaders are the TXs and the followers are the RXs. The leaders first choose to transmit the semantic concepts with the goal of transmitting as few bits as possible while ensuring the QoTE is close to one. 
This is followed by the RXs' decisions on whether to engage in reasoning, accept the received message, or take no action, to maximize their QoTE. 
We define the resulting Stackelberg game as $\mG = \left( \mK \cup \mJ, \mA_l\cup\mA_f, \mU_l \cup \mU_f \right)$, where index $l$ corresponds to the leader and $f$ to the followers. $\mA_l = \mA_{l,1} \times\cdots\times \mA_{l,K/2}$ and $\mA_{l,k}$ is defined as the strategy set of TX $k$. $\mA_f = \mA_{f,1} \times\cdots\times \mA_{f,K/2}$ and $\mA_{f,j}$ is the strategy set of RX $j$. We define $\mU_{l} = U_{l,1}\times\cdots\times\mU_{l,K/2}$, where $U_{l,k}: \mA_{l} \times \mA_f \rightarrow \mR$ is utility function of leader $k$. Similarly, we define $\mU_{f} = U_{f,1}\times\cdots\times\mU_{f,K/2}$, where $U_{f,j}: \mA_{l} \times \mA_f \rightarrow \mR$ is utility function of follower $j$. 

The communication decision of each TX is defined as the number of physical bits $a_{l,kr}$ allocated to a semantic concept $c_{kr}$. The number of physical bits communicated from any TX $k$ will be computed as  $\!\!\sum\limits_{c_{kr}\in\mC_{k}}\!\frac{1}{J}\sum\limits_{j \in \mJ}w_{kr}^{(j)} a_{l,kr} ,$
%
where the compression factor $w_{kr}^{(j)}$ is the same as the semantic  relevance factor $c_{kr}$ with respect to enhancing the reliability of RX $j$'s task $T_j$. The total number of bits that can be communicated across the network is limited by an upper bound $B$, due to the finite bandwidth constraint. Hence, we define TX $k$'s mixed strategy as the probability vector $\pi(\bma_{l,k}\mid\mC_k) \in \mA_{l,k}$, where the strategy set $\mA_{l,k}$ consists of $\pi(\bma_{l,k}\mid\mC_k)$, that satisfies the constraint:
$ \sum\limits_{k\in \mK}\sum\limits_{c_{kr}\in\mC_{k}}\mbE_{\pi}\left[\frac{1}{J}\sum\limits_{j \in \mJ}w_{kr}^{(j)} a_{l,kr}\right] \leq B,$
where the expectation is with respect to $\pi(\bma_{l,k}\mid \mC_k)$. Here, $\bma_{l,k}$ is a vector with $r^{\textrm{th}}$ element $a_{l,kr}$. 
The matrix of semantic relevance vectors $\bmW_j = \left[\bmw_{1}^{(j)},\cdots,\bmw_{\frac{K}{2}}^{(j)}\right]$ is defined as the preference matrix for RX $j$ and is unknown to the TXs.

For RX $j$, the computing decisions are defined by the vector $\bma_{f,j}$ of dimension $\frac{K}{2}\times 1$. Each scalar element ${a}_{f,kj}$ in $\bma_{f,j}$ represents a decision on whether to use the received information from TX $k$ (based on the received signal quality), perform reasoning computations at the local computing server or the CC server, or drop the packets. These decisions are captured by values $0$, $1$, $2$, and $3$, respectively. 
The strategy space of any RX $j$ is:
\vspace{-2mm}\beq
\begin{aligned}
    &\mA_{f,j} = \Bigg\{\pi(\bma_{f,j}\mid \widehat{\mC}) \mid \sum\limits_{k \in \mK} \pi(a_{f,kj}=1\mid \widehat{C}_k^{(j)}) F_{kr}^{(j)} \leq F_j^{\mathrm{max}}, 
\\    
 &\sum\limits_{k\in \mK}\sum\limits_{j \in \mJ, a_{f,kj}=2} \pi(a_{f,kj}=2\mid \widehat{C}_k^{(j)}) \sum\limits_{r=1}^D F_{kr}^{(j)} \leq F_0^{\mathrm{max}} \Bigg\}.
\end{aligned}
\label{eq_RX_strategy}
\vspace{-1mm}\eeq
Clearly, the strategy space of any RX is dependent on the other RX strategies via the shared computing resource constraints. For notational simplicity, we abbreviate the mixed strategy action probabilities for leaders and followers using the vectors $\boldsymbol{\pi}_{l,k}$ and $\boldsymbol{\pi}_{f,j}$, respectively. Next, we look at the utility functions of each user.

\vspace{-1mm}\subsection{Utility Functions}\vspace{-1mm}

First, we look at the utility function of each TX.  The average semantic reliability of the concepts communicated from TX $k$ can be captured as the average semantic surprise across the decodable links, $V_{l,k}(\{\bpi_{f,j}\}_{\forall j\in\mJ}) = -\sum\limits_{j\in \mJ}\pi(a_{f,kj}=0\mid \widehat{\mC}_k^{(j)})\mathbb{E}_p\left[\sum\limits_{r=1}^Dw_{kr}^{(j)}\log p(\widehat{c}_{kr}^{(j)}\mid c_{kr})\right]$, where $\mathbb{E}_p$ is the expectation using $p(\widehat{c}_{kr}^{(j)}\mid c_{kr})$. The \emph{Semantic surprise} $V_{l,k}$ quantifies the degree to which the content of the transmitted message is surprising to its recipients. $V_{l,k}$ depends on the multiple access channel distribution from all TXs to RX $j$ and it is captured using $p(\widehat{c}_{kr}^{(j)}\mid c_{kr})$.  
Each TX $k$ seeks to minimize the number of physical bits communicated, while simultaneously ensuring that the average semantic surprise across all users is below a threshold. Hence, the utility function of TX $k$ is dependent on the semantic relevance factors of each RX as well as the random channel distribution and is given by:
\vspace{-2mm}\beq
\vspace{-1mm}\begin{aligned}
&u_{l,k}(\bpi_{l,k},\bpi_{l,-k},\{\bpi_{f,j}\}_{\forall j \in \mJ},\{\bmw_k^{(j)}\}_{j\in\mJ}) \\ & = \alpha_1\underbrace{\frac{1}{J}\sum\limits_{j\in \mJ}\mbE_{\pi_{l,k}}\left[\boldsymbol{1}^T(\bmw_k^{(j)}\otimes\bma_{l,k})\right]}_{\textrm{avg. number of bits}} + \alpha_2 V_{l,k}(\{\bpi_{f,j}\}_{\forall j\in\mJ}),
\end{aligned}
\label{eq_TX_utility}
\vspace{-0mm}\eeq
where $\alpha_1+ \alpha_2=1$ and $ \bma_{l,k} \in \mA_{l,k}$. $\boldsymbol{1}$ is the vector of all ones and $\otimes$ represents the element-wise multiplication. Due the constraint on the total number of bits that can be communicated across all TXs, the strategy of TX $k$ depends on that of other TXs. This leads to the dependency of utility $u_{l,k}$ on $\bpi_{l,-k}$, which is the vector of strategies of all TXs except $k$. The  weights $\alpha_1$ and $\alpha_2$ represent a tradeoff between \emph{minimizing the number of physical bits communicated and maximizing the semantic reliability of task execution}. $\alpha_1=0$ represents the case when the emphasis is on maximizing the reliability of task execution at RX $j$, but results in inefficient usage of transmission resources. 
 The second term in \eqref{eq_TX_utility}, $V_{l,k}$, is contingent upon whether the RX relies on reasoning or the information communicated through bits. Consequently, the TX utility becomes dependent on the strategies adopted by the RX, transforming the scenario into a Stackelberg game. 

The utility function of each RX must capture the tradeoff between semantic concept reconstruction quality and the reasoning success probability. The semantic concept reconstruction quality of any RX $j$ is affected by the reliability of communication links from TXs and hence, the number of bits used to represent any semantic concept $c_{kr}$. This means that RX $j$ strategy depends on the TX strategies $\mA_{l,k}$. If the communication link quality does not allow the semantic concepts to be decoded, then $\mA_{f,j}$ depends on the availability of local and shared computing resources to perform reasoning in case. Hence, the strategies of the RXs are interdependent.  
The semantic concept reconstruction quality for the link from $k$,  $E(\mC_{k}^{(j)}, \widehat{\mC}_{k}^{(j)})$ is given by \eqref{eq_recon_error},
\begin{figure*}
\beq
\begin{aligned}
    &E(\mC_{k}^{(j)}, \widehat{\mC}_{k}^{(j)})  = \underbrace{\mathbb{E}_p(\sum\limits_{r=1}^D\pi(a_{f,kj} = 0\mid \widehat{\mC}_k^{(j)}) w_{kr}^{(j)}\abs{c_{kr} - \widehat{c}_{kr}^{(j)}}^2)}_{\textrm{Communication quality}}  + \underbrace{\mathbb{E}_p(\sum\limits_{r=1}^D \pi(a_{f,kj} = 1\mid \widehat{\mC}_k^{(j)}) w_{kr}^{(j)}\abs{c_{kr} - \widehat{c}_{kr}^{(j)}}^2)}_{\textrm{Local reasoning accuracy}}  \\ & + \underbrace{\mathbb{E}_p(\sum\limits_{r=1}^D\pi(a_{f,kj} = 2\mid \widehat{\mC}_k^{(j)}) w_{kr}^{(j)} \abs{c_{kr} - \widehat{c}_{kr}^{(j)}}^2)}_{\textrm{CC server reasoning accuracy}} +  \underbrace{\mathbb{E}_p(\sum\limits_{r=1}^D\pi(a_{f,kj} = 3\mid \widehat{\mC}_k^{(j)}) w_{kr}^{(j)} P_e)}_{\textrm{Penalty when data is dropped}},
\end{aligned}
\label{eq_recon_error}
\eeq\vspace{-5mm}
\end{figure*}
\begin{figure*}
\beq
\begin{aligned}
\tau_{kj} &= \sum\limits_{r=1}^D \underbrace{p(\widehat{c}_{kr}^{(j)}\mid {c}_{kr}^{(j)})\pi(\bma_{f,kj} = 1\mid \widehat{\mC}_k^{(j)}) \frac{F_{kr}^{(j)}}{F_{j}^{\mathrm{max}}}}_{\textrm{Local computing delay}} +\sum\limits_{r=1}^D  \underbrace{p(\widehat{c}_{kr}^{(j)}\mid {c}_{kr}^{(j)})\pi(\bma_{f,kj} = 2\mid \widehat{\mC}_k^{(j)}) \left(\frac{w_{kr}^{(j)}a_{l,kr}}{R_j} + \frac{F_{kr}^{(j)}}{d_{j}}\right)}_{\textrm{CC server computing delay}}.
\end{aligned}
\label{eq_utility_delay}
\eeq\vspace{-2mm}
\vspace{-1mm}\end{figure*}
where $P_e$ is the penalty applied when the data is dropped due to unreliable communication link and insufficient computing resources for reasoning. However, the cost associated with dropping the data depends on the sum of the semantic relevance across all concepts transmitted from TX $k$. If this sum of semantic relevance is zero, then there is no cost associated with discarding the data. This is where a semantic-aware computing and communication resource allocation can help to efficiently use the available resources. In a general sense, we can write the first summation term in \eqref{eq_recon_error} as a function of the channel distribution and the number of bits transmitted, i.e., $\abs{c_{kr} -\widehat{c}_{kr}^{(j)}}^2 = f(p\left(\widehat{c}_{kr}^{(j)}\mid c_{kr}),w_{kr}^{(j)}a_{l,kr}\right)$. For a zero mean Gaussian distribution $p\left(\widehat{c}_{kr}^{(j)}\mid c_{kr}\right)$ with variance $\sigma^2$, we can obtain the minimum distortion as \cite{CoverThomas1991} $f(p\left(\widehat{c}_{kr}^{(j)}\mid c_{kr}),w_{kr}^{(j)}a_{l,kr}\right) = \sigma^2 2^{-2\frac{w_{kr}^{(j)}a_{l,kr}}{T}}$. Finally, we observe that \eqref{eq_recon_error} is a function of TX strategies $\bpi_{l,k}$. The second term in \eqref{eq_recon_error} can be obtained by substituting \eqref{eq_interventions} for $\widehat{c}_{kr}^{(j)}$.

The third component of the utility function is the average delay \eqref{eq_utility_delay}  at RX $j$ in computing the semantic concepts relevant to $T_j$.
Here, we disregard the delay related to communication from the TXs, assuming they are high-rate links with insignificant delays compared to reasoning computations. We also assume no re-transmissions, as we rely on computing capabilities to predict and correct erroneous transmissions through reasoning. This is an advantage of employing semantics-aware resource allocation by allowing flexibility in avoiding re-transmissions and reducing error correction overheads. If the reasoning computations are not received within a stipulated time frame ($\leq \tau^{\mathrm{max}}$), we consider that the packets are dropped, affecting the QoTE defined as $\mQ_j$. 
Further, we write the utility function of RX $j$ in \eqref{eq_QoTE} as the inverse of QoTE,
\begin{figure*}
\vspace{-2mm}\beq
\begin{aligned}
&u_{f,j}(\{\bpi_{l,k}\}_{\forall k\in\mK },\bpi_{f,j},\bpi_{f,-j}) = &  \sum\limits_{k\in \mK} \left[P(\tau_{kj} \leq \tau^{\mathrm{max}})E(\mC_{k}^{(j)}, \widehat{\mC}_{k}^{(j)}) + (1-P(\tau_{kj} \leq \tau^{\mathrm{max}}))P_q\right]
\label{eq_QoTE}
\end{aligned}
\vspace{-6mm}\eeq
\end{figure*}
where $P_q$ is the penalty for reasoning failures.
The reasoning computations delay $\tau_{kj}$ depends on the strategies of other RXs $\bpi_{f,-j}$, given the shared CC resources. Computing $P(\tau_{kj} \leq \tau^{\mathrm{max}})$ is not trivial. Hence, we next, derive an upper bound for the reasoning success probability.
\vspace{-2mm}\begin{lemma}
\label{Lemma1}
    The reasoning success probability $P(\tau_{kj} \leq \tau^{\mathrm{max}})$ for Gaussian distributions $p(\widehat{c}_{kr}^{(j)}\mid c_{kr})$ is upper bounded by $e^{-\frac{x_{kj}^2}{2}}$, where   
     $x_{kj}\! = \! \frac{N_0B(\!2^{\frac{\beta_{kj}(2)}{B(\tau^{\mathrm{max}} - \beta_{kj}(1))}} \!-\! 1)}{P_i\sigma}$.
\end{lemma}\vspace{-2mm}
\begin{IEEEproof}
We define $\beta_{kj}(2) = \sum\limits_{r=1}^D  p(\widehat{c}_{kr}^{(j)}\mid {c}_{kr}^{(j)})\pi(a_{f,kj}= 2\mid \widehat{\mC}_k^{(j)})w_{kr}^{(j)}a_{l,kr}$ and $\beta_{kj}(1) = \sum\limits_{r=1}^D p(\widehat{c}_{kr}^{(j)}\mid {c}_{kr}^{(j)})\pi(a_{f,kj}= 1\mid \widehat{\mC}_k^{(j)}) \frac{F_{kr}^{(j)}}{F_{j}^{\mathrm{max}}} + \sum\limits_{r=1}^D  p(\widehat{c}_{kr}^{(j)}\mid {c}_{kr}^{(j)})\pi(a_{f,kj} = 2\mid \widehat{\mC}_k^{(j)}) \frac{F_{kr}^{(j)}}{d_{j}} $. Further, substituting for $\tau_{kj}$ from \eqref{eq_utility_delay} and $R_j$, we obtain,
   \beq
   \begin{aligned}\small
  P(\tau_{kj} \leq \tau^{\mathrm{max}}) &=  P\left(\abs{h_i} \geq \frac{N_0B\left(2^{\frac{\beta_{kj}(2)}{B(\tau^{\mathrm{max}} - \beta_{kj}(1))}} - 1\right)}{P_i}\right). 
  \end{aligned}
   \eeq
   Further, the bound in Lemma~\ref{Lemma1} follows from using the $Q$-function of $h_i \sim \mN(0,\sigma^2)$. 
\end{IEEEproof}

\vspace{-2mm}\subsection{Hypergame Theory for Handling Misperceptions}\vspace{-1mm}

The utility functions $u_{l,k}$ and $u_{f,j}$ show intricate inter-dependencies among user strategies and preferences.
Here, the TXs are unaware of the semantic relevance (that somewhat represent the preferences of each RX) of each extracted concept for the task $T_j$ of RX $j$. 
This misperception poses a challenge for the TXs in computing an optimal strategy, as they strive to minimize the amount of communicated bits while ensuring the QoTE is maximized. We assume that TX $k$ holds a belief $\widehat{w}_{kr}^{(j)}$ about the preferences of any RX $j$.  
Furthermore, each TX $k$ lacks information regarding the strategies adopted by other TXs. It is possible that there are common semantic concepts extracted from observations of any two TXs, due to correlation between their observations. Hence, communicating redundant information results in inefficient usage of communication resources. To avoid transmitting redundant information, TXs must be aware of each other’s strategies.
Similarly, RXs are unware of the strategies $\bpi_{f,i}$ of any other RX $i \in \mJ$ and their channel qualities, in order to decide whether the offloading the reasoning to CC server results in data being dropped or not, in scenarios where the communication link is not reliable and local computing resources are not sufficient for reasoning. 
 A promising approach here is to compute close to optimal strategies by incorporating these misperceptions is by using the framework of \emph{hypergame theory} \cite{KovachGT2015}. Hypergame theory is a promising approach here since it allows each user to have its own perception about other user strategies and update those perceptions based on the game outcomes $\mS$, to tune each user strategies. Next, we define a second level hypergame.

\vspace{-1mm}\begin{definition}[\emph{Second level hypergame}]
    A second level hypergame involves the situation where each user is aware that there is a hypergame going on. It is defined as a set of first level hypergames played by each user $\mH^2 = ( \mH_{s,i},\forall i \in \{\mK,\mJ\}, \forall s \in \{l,f\})$, where each $\mH_{s,i} = (\mG_{i1},\cdots,\bmG_{ii},\cdots,\mG_{iK}), \forall s \in \{l,f\}$, where $\mG_{iq}$ represents user $q$'s game under $i$'s perception. $\mG_{ii}$ is user $i$'s game under its own perception. We define the set of parameters that are perceived by any user $j$ about user $i$ using the variable $\btheta_{ij} \in \bTheta$. For any follower $j$, $\btheta_{kj} = \{\widehat{\bpi}_{l,k}^{(j)}\}, \forall k\in \mK$, $\btheta_{ij} = \{\widehat{\pi}_{f,i}^{(j)}, \forall k \in \mK\}, \forall i\in \mJ$. For any leader, $\btheta_{jk} = \{\widehat{\pi}_{f,j}^{(k)},\forall j \in \mK, \widehat{\mW}_{j}^{(k)}\}$. $\mU^i = \times_{s,j} u_{s,j}^{i}, s\in \{l,f\}$ is a profile of utility functions perceived by user $i$, where $u_{s,j}^i$  is user $j$'s utility function perceived by user $i$.
\end{definition}

\vspace{-2mm}\section{Equilibrium Analysis}\vspace{-2mm}

Next, we look at the solutions for the TX and RX strategies of the resulting hypergame using HSE under misperception.
Consider that TX $k$ has a perception of the strategies of other TXs as $\widehat{\bpi}_{l,-k}^{(k)}$ and semantic relevance as $\widehat{\mW}_j^{(k)}$. In this regard, TX $k$'s utility under misperception can be written as:
\vspace{-2mm}\beq
\begin{aligned}
&u_{l,k}(\bpi_{l,k}^{(k)},\widehat{\bpi}_{l,-k}^{(k)},\{\widehat{\mW}_j^{(k)}\}_{j\in\mJ},\{\bpih_{f,j}^{(k)}\}_{\forall j \in \mJ})  =\\ & \alpha_1\frac{1}{J}\sum\limits_{j\in \mJ}\mbE_{\pi_{l,k}}\left[\boldsymbol{1}^T(\widehat{\bmw}_k^{(j)}\otimes{\bma}_{l,k})\right] + \alpha_2 V_{l,k}(\{\widehat{\bpi}_{f,j}^{(k)}\}_{\forall j\in\mJ}),
\end{aligned}\vspace{-1mm}
\eeq
where $\widehat{\mW}_j^{(k)}$ is the set of all semantic relevance factors for RX $j$ as perceived by TX $k$.
Correspondingly, the follower $j$'s best response (as perceived by $k$) to the
leaders strategies $\bpi_{l,k}^{\ast}$ under the prejudiced beliefs $\btheta_{jk}, \forall j \in \mJ$ can be written as:
\beq
\begin{aligned}
    &r_{f,j}(\{\bpi_{l,k}^{\ast}\}_{\forall k \in \mK})  = \max_{\bpi_{f,j} \in \mA_{f,j}} u_{f,j}(\{\bpi_{l,k}^{\ast}\}_{\forall k \in \mK},\bpih_{f,j}^{(k)},\bpih_{f,-j}^{(k)}).
\end{aligned}
\vspace{-2mm}\eeq
However, if the users update the perception $\btheta_{ij}$ based on the outcomes $\mS$ of the hypergame, analyzing the equilibria requires defining the concept of HSE. Here, we define $\mN(\mH_{s,i}), \forall s\in \{l,f\}, i\in \{\mK,\mJ\}$ as the mixed strategy Nash equilibrium (NE) of the game played by user $i$.
\begin{definition}[Hyper Stackelberg Equilibrium, HSE]
\label{HSE}
   Expanding the results in \cite{GharesifardCDC2010} for a multi-leader and multi-follower static Stackelberg hypergame, a strategy profile $\pi(\bma_{s,i}\mid\mC) \in \mA_{s,i}, s\in \{l,f\}, i\in \{\mK,\mJ\}$ is called an HSE of a hypergame iff $\forall i \in \{\mK,\mJ\}$, $\pi(\bma_{s,i}\mid \mC) \in \mN(\mH_{s,i})$. 
\end{definition}
Definition~\ref{HSE} means that user strategies converge to an HSE when they reach the Nash equilibrium (NE) solution in their subjective games, based on a certain perception of unknown information about others. HSE can be analytically defined through a two-stage process. Initially, we establish the equilibrium for the leaders' game (which is a simultaneous move) by analyzing the best responses from the followers: 
\vspace{-1mm}\beq
\begin{aligned}
    &\bpi_{l,k}^{\ast} \in \argmin\limits_{\substack{\bpi_{l,k}\in \mA_{l,k}}} u_{l,k}(\bpi_{l,k},\btheta_{jk}, \forall j\in \{\mK,\mJ\}), \forall k \in \mK, \\ 
    &\widehat{\bpi}_{f,j}^{(k)\,\ast} \in  \argmin\limits_{\bpih_{f,j}^{(k)}\in \mA_{f,j}}\widehat{u}_{f,j}(\bpih_{f,j}^{(k)},\btheta_{jk}, \forall k\in \{\mK,\mJ\}), \forall j \in \mJ.
\end{aligned}
\label{eq_HSE_leader}
\eeq\vspace{-1mm}
Second, we establish the equilibrium for the followers' game:
\beq\vspace{-1mm}
\begin{aligned}
    &\widehat{\bpi}_{l,k}^{(j)\,\ast} \in \argmin\limits_{\widehat{\bpi}_{l,k}^{(j)}\in \mA_{l,k}} \widehat{u}_{l,k}(\bpih_{l,k}^{(j)},\btheta_{ij}, \forall i\in \{\mK,\mJ\}), \forall k \in \mK, \\ 
    &\bpi_{f,j}^{\ast} \in\argmin\limits_{\bpi_{f,j}\in \mA_{f,j}}u_{f,j}(\bpi_{f,j},\btheta_{kj}, \forall k\in \{\mK,\mJ\}), \forall j \in \mJ.
\end{aligned}
\label{eq_HSE_follower}
\vspace{-0mm}\eeq
Following the actions taken by TXs and RXs as per \eqref{eq_HSE_leader} and \eqref{eq_HSE_follower}, the users adjust their perceptions (details of the perception updates are discussed in Section~\ref{Perception}) in response to the outcomes. 
We define the set of HSEs as $\mN^H = \times_{s,i} \mN(\mH_{s,i})$. When the users does not update the perception based on the game outcomes or is not aware of any misperception, we call the corresponding equilibrium as the misperception strong Stackelberg equilibrium (MSSE) \cite{ChengTIFS2022}. Given that the users update their perception $\btheta_{ij}$ based on the game outcome, the following inequalities hold: $u_{l,k}^{(HSE)} \leq u_{l,k}^{(MSSE)}, \forall k \in \mK$ and $u_{f,j}^{(HSE)} \leq u_{f,j}^{(MSSE)}, \forall j \in \mJ$. However, it's important to note that HSE does not inherently ensure the stability of the TX and RX strategies. This implies that for any HSE strategy for user $k$, $\{\bpi_{l,k}^{\ast},\bpih_{s,j}^{(k)\,\ast},\forall s\in \{l,f\}, \forall j \in \{\mK,\mJ\} \}$, the optimal response from the followers may not align with their respective HSE strategy. This means that $\bpi_{f,j}^{\ast}$ need not be equal to $\bpih_{f,j}^{(k)\,\ast}$. Similarly, the same disparity between optimal responses and HSE strategies can also occur for the followers. 

\vspace{-2mm}\subsection{Proposed alternating minimization solution}

To solve \eqref{eq_HSE_leader}, we follow an alternating minimization (AM) approach, wherein we update the TX strategies and the perceptions alternatively. 
\subsubsection{TX and RX strategy updates}
To derive the TX strategies for a given $\btheta_{jk}$, we observe that the utility $u_{l,k}$ is convex with respect to TX actions $a_{l,kr}$ and the constraints are linear (and hence convex) and compact. Hence, we write the resulting Lagrangian as \eqref{eq_Lagrangian_Leader} and derive the strategies as below.
\begin{figure*}\vspace{-0mm}\beq
\begin{aligned}
\mL_k &= u_{l,k}(\bpi_{l,k},\widehat{\bpi}_{l,-k}^{(k)},\{\widehat{\mW}_j^{(k)},\widehat{\bpi}_{f,j}^{(k)}\}_{j\in\mJ}) + \lambda \mbE_{\pi}\left(\sum\limits_{k\in \mK}\sum\limits_{c_{kr}\in\mC_{k}}\frac{1}{J}\sum\limits_{j \in \mJ}\widehat{w}_{kr}^{(j)} a_{l,kr} \leq B\right).
\label{eq_Lagrangian_Leader}
\end{aligned}
\eeq\vspace{-6mm}\end{figure*}
\vspace{-2mm}\begin{lemma}
\label{lemma2}
    For a given set of semantic concepts $\mC_k$, semantic relevance factors $w_{kr}^{(j)}$, and TX strategies for all $i\neq k$, the TX strategy for user $k$ can be obtained as \eqref{eq_Wkj},
    where $W_{kr}^{(j)} = w_{kr}^{(j)}a_{l,kr}$ and $\lambda^{\ast}$ is computed using bisection \eqref{eq_bisection}.\vspace{-2mm}
\end{lemma}
\begin{IEEEproof}
    The proof is provided in Appendix~\ref{proof_lemma2}.
\end{IEEEproof}
Lemma~\ref{lemma2} implies that the number of bits allocated to any semantic concept $c_{kr}$ is inversely proportional to both the channel quality indicated by $p(\widehat{c}_{kr}^{(j)}\mid c_{kr}^{(j)})$ and the probability that the reconstructed concept is not dropped for task execution. Intuitively, this suggests that on average (across all links), if a particular semantic concept is either decodable or can be reasoned using computing powers at the RX or CC side, then such concepts can be encoded with fewer bits.
\vspace{-2mm}\begin{lemma}
\label{lemma3}
    For a given set of semantic concepts $\mC_k$, semantic relevance factors $w_{kr}^{(j)}$, and RX strategies for all $i\neq j$, the RX strategy for user $j$ can be obtained as follows:
    \vspace{-2mm}\beq
    \pi(a_{f,kj}=0\mid \widehat{C}_k) = p(\abs{c_{kr}-\widehat{c}_{kr}}^2 \geq \delta),
    \eeq
    $\pi(a_{f,kj}=1\mid \widehat{C}_k)$ and $\pi(a_{f,kj}=2\mid \widehat{C}_k)$ as \eqref{eq_pi_akj_equi}, and
    \vspace{-3mm}\beq
    \pi(a_{f,kj}=3\mid \widehat{C}_k) = 1-\sum\limits_{d=0}^2\pi(a_{f,kj}=d\mid \widehat{C}_k).\label{eq_pi_akj_3}
    \vspace{-1mm}\eeq\vspace{-2mm}
 \end{lemma}
 \vspace{-2mm}\begin{IEEEproof}
     The proof is provided in Appendix B of the arxiv version of the paper.
 \end{IEEEproof}
\eqref{eq_pi_akj_equi} means that reasoning computation decisions depend inversely on the reasoning success probability. Moreover, the decision on where to perform the computations depends on how the computing resources at the CC server are shared among the RXs, which requires the knowledge of other RX strategies. $\mL_k$ is convex with respect to $\pi_{l,k}$. However, $\mL_j, \forall j \in \mJ$ is concave with respect to the RX strategies and $\mW_{j}$ and are solved by convex approximations of $\mL_j$, as discussed in Lemma~\ref{lemma3}. Hence, the proposed Stackelberg hypergame strategies converge to a local HSE. Next, we propose a swap learning method \cite{GharesifardCDC2010} to update the perceptions of the TX and RX based on the outcomes of user actions, as described in Lemma~\ref{lemma2} and Lemma~\ref{lemma3}.
\vspace{-0mm}\subsubsection{Perception updates for two-user system}
\label{Perception}

For simplicity, we consider the evolution of perception as our game progresses for a single TX (player A) and single RX system (player B). We define the hypergame as $\mH_A = \left(\mG_{AA},\mG_{BA}\right)$ and $\mH_B = \left(\mG_{BB},\mG_{AB}\right)$. 
Following \cite{GharesifardCDC2010}, we consider that there is an H-diagraph associated to player
$A$’s hypergame, where the nodes in the graph $\bms\in \mS$ are labeled with $\left(u_{l,A}(\bms,\mG_{AA}),u_{f,B}(\bms,\mG_{BA})\right), \bms\in \mS$. There exists an edge from the outcome $\bms_1$ to $\bms_2$, iff there exists an improvement $\bms_2$ from $\bms_1$ for player $A$ in the
game $\mG_{AA}$ and for
player $B$ in the game $\mG_{BA}$ for which there exists no perceived sanction
of $B$ in the game $\mG_{BA}$ or $A$ in the game $\mG_{AA}$, respectively. For this two-player hypergame, we can further define the misperception function as $\mM_{BA}:\mS\times \mS\rightarrow \mR_{\geq 0}$, given by
\vspace{-2mm}\beq
\mM_{BA}(\btheta_{BA}) = \sum\limits_{i=1}^N \abs{u_{f,B}(\bms_i,\mG_{BB}) - u_{f,B}(\bms_i,\mG_{BA})},
\label{eq_misperception_func}
\vspace{-2mm}\eeq
where $N$ is the number of possible outcomes.
Similarly, we can define $\mM_{AB}(\btheta_{AB}) = \sum\limits_{i=1}^N \abs{u_{l,A}(\bms_i,\mG_{AA}) - u_{l,A}(\bms_i,\mG_{AB})}$. \emph{Misperception function} is a measure to quantify the error in perception about opponent's utility function. Furthermore, swap learning can be described as follows. Suppose player $B$ adopts a strategy that alters the game outcome from $\bms_i$ to $\bms_j$, reducing the utility function value for player $B$. If, in this scenario, player $A$'s optimal strategy remains unchanged, player $A$ updates their perception as $\btheta_{BA} = \mT_{\bms_i \rightarrow \bms_j}(\btheta_{BA})$. Here, $\mT_{\bms_i \rightarrow \bms_j}$ represents the projection operator of player $B$'s strategy and preferences corresponding to the transition from $\bms_i$ to $\bms_j$. The updates $\mT_{\bms_i \rightarrow \bms_j}$ can be implemented using gradient descent (with step size $\eta$), as follows:
\vspace{-1mm}\beq
\btheta_{BA}^{(t+1)} = \btheta_{BA}^{(t)}-\eta \nabla \mM_{BA}(\btheta_{BA})|_{\btheta_{BA}^{(t)}}.
\label{eq_gradientdescent}
\vspace{-1mm}\eeq
Similarly, we define the perception updates for $\btheta_{AB}$. To extend swap learning to more than two users, we define the misperception between each pair of users $A$ and $B$, where $A, B \in {\mK, \mJ}$. Their perceptions of each other are updated using the gradient descent approach described in \eqref{eq_gradientdescent}.
\vspace{-1mm}\subsection{Convergence of the proposed solution}
Further, we look at how the misperception function is affected by the swap learning update.
\vspace{-2mm}\begin{lemma}
\label{lemma_monotonicity}
    For the two player hypergame, suppose player $B$ takes a rational action
such that the outcome of the hypergame changes from $\bms_i$ to $\bms_j$, with $\pi_A(\bma_{A}\mid \mC)$ remains intact. Let, the resulting misperception function be defined as $\mM^{\prime}_{BA}$. Then, the misperception function always decreases under swap learning, i.e.,
$\mM_{BA}^{\prime}(\btheta_{BA}^{\prime}) \leq \mM_{BA}(\btheta_{BA})$, since the updated beliefs reduces the value of the utility functions.
\end{lemma}
\begin{IEEEproof}
    Since player $B$ follows the NE strategies as derived in Lemma~\ref{lemma2} and~\ref{lemma3}, $u_{f,B}(\bms_j,\mG_{BB}) \leq u_{f,B}(\bms_i,\mG_{BB})$. If player $A$ updates its perception about $B$ based on \eqref{eq_gradientdescent}, then the $u_{f,A}(\bms_j,\mG_{BA})$ should also monotonically decrease. However, due to the misperceptions, $u_{f, A}(\bms_j,\mG_{BA})$ will still be greater than $u_{f,B}(\bms_j,\mG_{BB})$, but the gap will reduce due to the updated perception. Hence, the misperception function monotonically decreases, $\mM_{BA}^{\prime}(\btheta_{BA}^{\prime}) \leq \mM_{BA}(\btheta_{BA})$.
\end{IEEEproof}
\begin{lemma}
\label{lemma_conv_mis}
    Under swap learning, the evolutions of the perceptions converge to local equilibrium solutions of $\mM_{BA}(\btheta_{BA})$ and $\mM_{AB}(\btheta_{AB})$, defined as $\btheta_{BA}^{\ast}$ and $\btheta_{AB}^{\ast}$, respectively.
\end{lemma}
\begin{IEEEproof}
    The proof directly follows from the monotonically decreasing nature of the misperception function under swap learning as mentioned in Lemma~\ref{lemma_monotonicity}. For monotonically decreasing misperception function, the gradient descent updates \eqref{eq_gradientdescent} converge to its local minimum.
\end{IEEEproof}
Generally, the ultimate misperception value in Lemma~\ref{lemma_conv_mis} may not be zero. This characteristic is typical in hypergames with outcome sets of large cardinality. Such hypergame strategies might result in an equilibrium where none of the players are inclined to alter their strategies further, even though certain portions of the outcome set remain unexplored.
 \begin{theorem}
 \label{Theorem1}
     The TX and RX strategies derived in Lemma~\ref{lemma2} and Lemma~\ref{lemma3}, respectively, converge to a local HSE for the second-level hypergame $\mH^2(\bTheta)$ when players update their perceptions using swap learning.
 \end{theorem}
\vspace{-2mm} \begin{IEEEproof}
 The proof is provided in Appendix~\ref{proof_theorem1}.
     \end{IEEEproof}
    Next, we numerically evaluate the local HSE solutions of the proposed TX and RX strategies and examine the tradeoff between communication and computing.
\begin{figure}[t]
\centerline{\includegraphics[width=3.3in,height=1.2in]{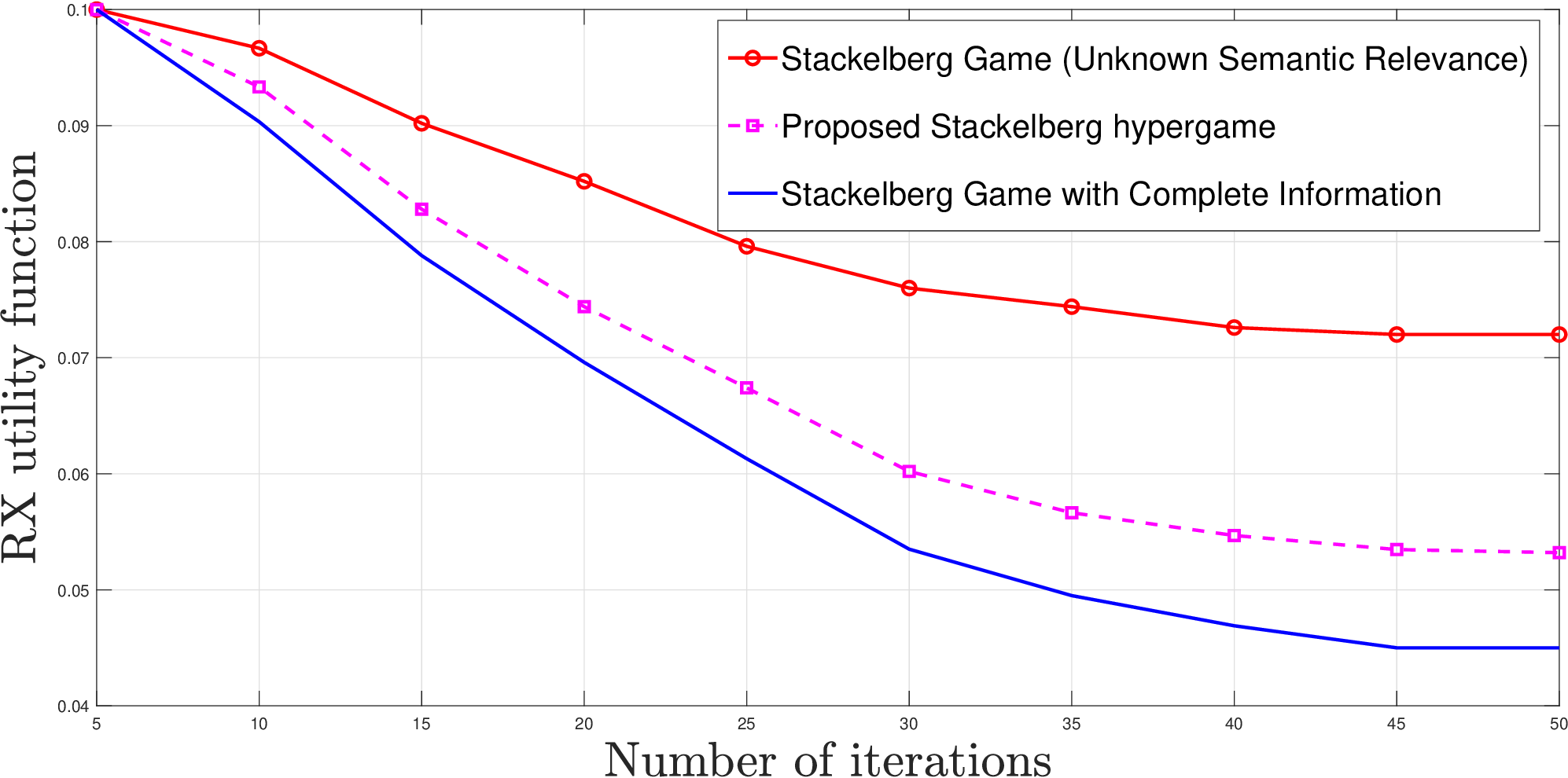}}\vspace{-1.5mm}
\caption{\small RX utility as a function of the number of game iterations.}
\label{RxUtility}\vspace{-0mm}
\vspace{-5mm}
\end{figure}
\begin{figure}[t]
\centerline{\includegraphics[width=3.3in,height=1.2in]{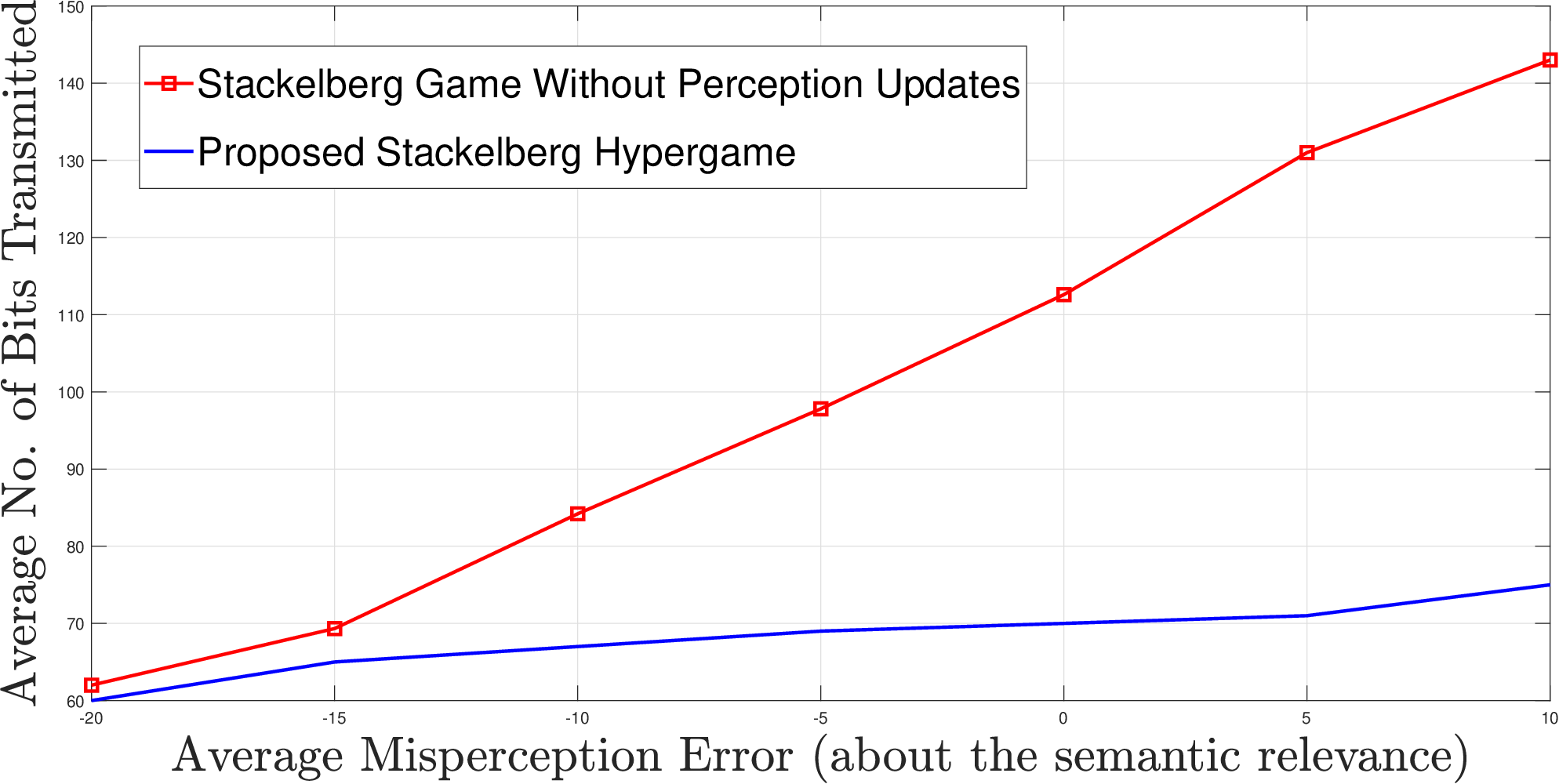}}\vspace{-1.5mm}
\caption{\small Number of bits transmitted vs function of Tx perception error.}
\label{TxMisperception}\vspace{-0mm}
\vspace{-5mm}
\end{figure}
\begin{figure*}
\vspace{-0mm}\beq
\begin{aligned}
    &\pi(a_{l,kr} \mid \mC_k) = \frac{-\log_2\left(\left(\frac{\alpha_1}{J} + \lambda\right)\frac{T}{\alpha_2\sigma^2}\right) \!-\! \dsum\limits_{d\neq r}\pi(a_{l,kd} \mid \mC_k)\frac{W_{kd}^{(j)}}{T}\!-\!\log2\left[ p(\widehat{c}_{kr}^{(j)}\mid \widehat{c}_{kr}^{(j)})\left[1-\pi(a_{f,kj} = 3\mid \widehat{\mC}_k^{(j)})   \right] \right]}{2W_{kr}^{(j)}T^{-1} }
\label{eq_Wkj}
\end{aligned}
\vspace{-4mm}\eeq\vspace{-2mm}
\end{figure*}
\begin{figure*}
     \vspace{-2mm}\beq
     \begin{aligned}
     &\pi(a_{f,kj} = d\mid \widehat{\mC}_k^{(j)}) = \frac{e^{-\frac{x_{kj}^2}{2}}\sum\limits_{r=1}^D p(\widehat{c}_{kr}^{(j)}\mid {c}_{kr}^{(j)}) w_{kr}^{(j)}\abs{c_{kr} - \widehat{c}_{kr}^{(j)}}^2 + \gamma_{kj} F_{kr}^{(j)}}{x_{kj}^{\prime}(\beta_{kj}(d)^0)x_{kj}e^{-\frac{x_{kj}^2}{2}}\left(\sum\limits_{r=1}^D p(\widehat{c}_{kr}^{(j)}\mid {c}_{kr}^{(j)})\frac{F_{kr}^{(j)}}{F_{j}^{\mathrm{max}}}\right)\sum\limits_{r=1}^D p(\widehat{c}_{kr}^{(j)}\mid {c}_{kr}^{(j)}) w_{kr}^{(j)} \abs{c_{kr} - \widehat{c}_{kr}^{(j)}}^2}, d\in \{1,2\}
     \label{eq_pi_akj_equi}
     \end{aligned}
     \eeq\vspace{-4mm}
     \end{figure*}
\begin{figure}[h]
\vspace{-1mm}
\centerline{\includegraphics[width=3.3in,height=1.4in]{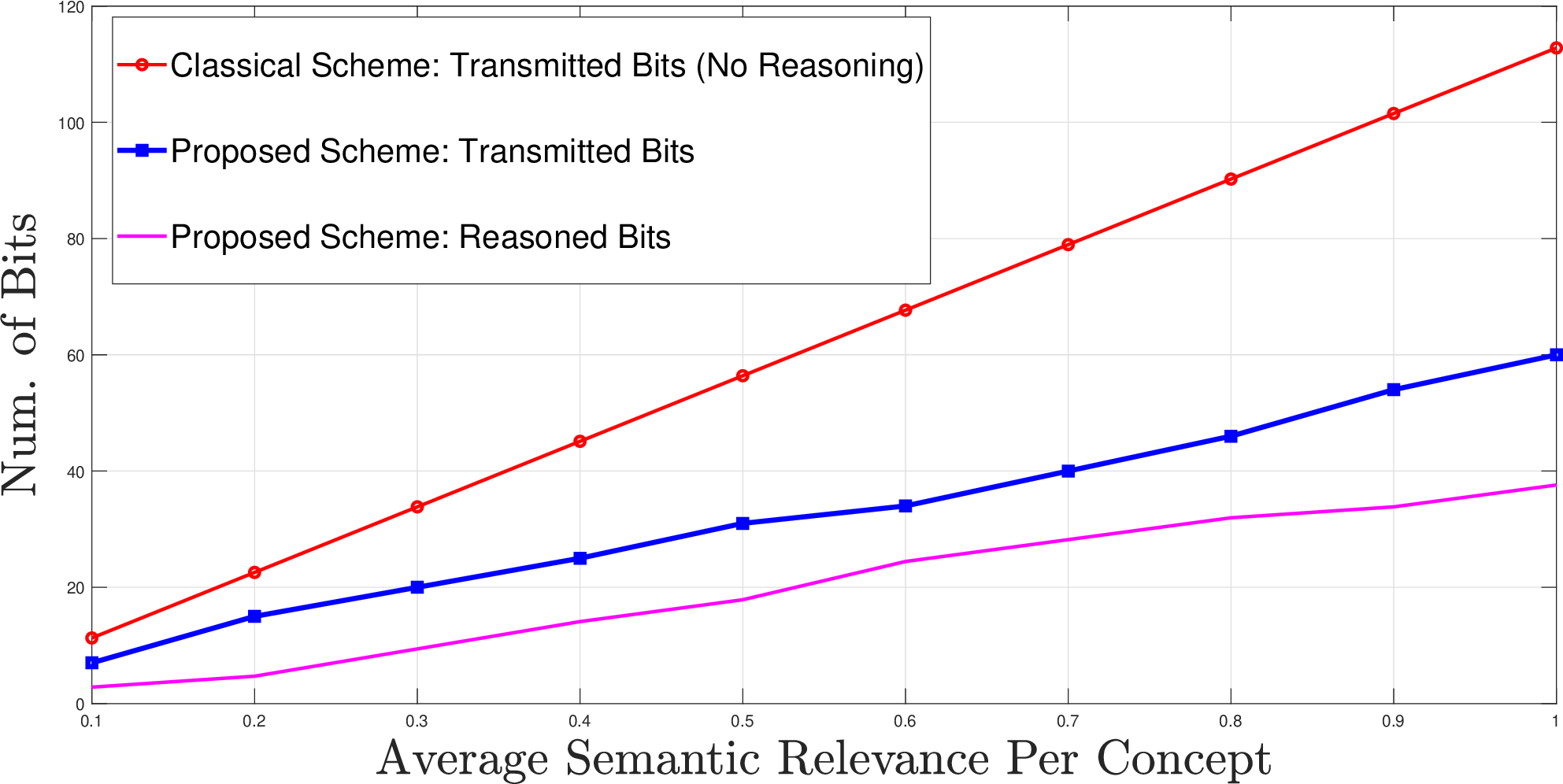}}\vspace{-2mm}
\caption{\small Number of bits communicated vs semantic relevance, for a fixed QoTE and $\tau^{\mathrm{max}}$.}
\label{CommReasoning}\vspace{-2mm}
\vspace{-6mm}
\end{figure}

\vspace{-0mm}\section{Simulation Results and Analysis}\vspace{-1mm}

We conducted extensive simulations to demonstrate the efficiency in terms of the QoTE as well as the transmission efficiency of the proposed Stackelberg hypergame based resource allocation. The simulation setup considered involves $2$ TXs and $2$ RXs. Semantic concepts are drawn from a Gaussian distribution with a distinct mean and variance equal to $1$. The mean values corresponding to distinct semantic concepts are drawn uniformly random from $[0,10]$. Further, a semantic relevance $w_{kr}^{(j)}$ of the concepts with respect to any user $j$'s task is exponential drawn from $c^{(i-1)}, i=1,\cdots,\abs{\mC}$. The value of $c$ determines the number of relevant semantic concepts for any user. We compared the proposed method with naive schemes based on hypergames and that does not incorporate the semantic relevance factor for TX strategies. 

Fig.~\ref{RxUtility} shows the converged RX utility values for the proposed scheme when the user strategies reach a local HSE. Moreover, the proposed scheme narrows the gap (error around $18\%$) in QoTE to Stackelberg game under complete information. The gap in performance compared to complete knowledge can be attributed to the fact that the proposed scheme converges to a local equilibrium solution.  Furthermore, the proposed scheme outperforms a Stackelberg game that does not integrate other user perceptions into their strategies.

Fig.~\ref{TxMisperception} illustrates that disregarding misperceptions regarding RX semantic relevance factors leads to significant inefficiency in transmission, as evidenced by the performance of the naive scheme. 
Moreover, the proposed scheme shows an improvement of nearly $52\%$ compared to the Stackelberg game scheme, which assumes that its knowledge (misperceived) of RX semantic relevance is perfect.

Figure~\ref{CommReasoning} shows that an SC system based on a Stackelberg hypergame outperforms alternative approaches in terms of resource allocation for both communication and computing resources. Specifically, the results indicate that the proposed scheme significantly reduces the number of physical bits communicated compared to a classical Stackelberg game approach that does not consider reasoning capabilities. This improvement is attributed to the incorporation of reasoning capabilities at the RX node and the consideration of semantic relevance for transmit resource allocation within the SC system.

\vspace{-3mm}\section{Conclusion}
\vspace{-2mm}

In this paper, we have introduced a novel decentralized resource allocation framework for multi-user SC systems using the Stackelberg hypergame theory. We formulated the HSE to jointly optimize transmission and reasoning strategies, aiming to minimize the number of bits communicated and enhance the QoTE at the RXs. Our simulations showed significant gains in the transmission resource usage while maintaining a high QoTE at the RXs compared to state-of-the-art resource allocation schemes that do not have a perception component about the incomplete information at the communicating nodes.

\vspace{-1mm}\appendices 

\vspace{-2mm}\section{Proof of Lemma~\ref{lemma2}}
\label{proof_lemma2}\vspace{-2mm}
For simplicity of analysis, we define $W_{kr}^{(j)} = w_{kr}^{(j)}a_{l,kr}$ as the number of encoded bits for concept $c_{kr}$ and intended to be transmitted to RX $j$. Taking the derivative of the Lagrangian \eqref{eq_Lagrangian_Leader} with respect to $\pi(a_{l,kr} \mid \mC_k)$ as $\frac{\partial \mL_k}{\partial \mbE_{\pi_{l,k}}W_{kr}^{(j)}}\frac{\partial \mbE_{\pi_{l,k}}W_{kr}^{(j)}}{\partial \pi(a_{l,kr} \mid \mC_k)}$, we obtain,
    \beq
    \begin{aligned}
    &\Bigg( \Bigg.\frac{\alpha_1}{J}   -\frac{\alpha_2\sigma^2}{T} p(\widehat{c}_{kr}^{(j)}\mid \widehat{c}_{kr}^{(j)})\left[ \left[\pi(a_{f,kj} = 0\mid \widehat{\mC}_k^{(j)}) 2^{-2\frac{\mbE_{\pi_{l,k}}W_{kr}^{(j)}}{T}}\right]\right. \\ & + \left. \left[ \pi(a_{f,kj} = 1\mid \widehat{\mC}_k^{(j)}) 2^{-2\frac{\mbE_{\pi_{l,k}}W_{kr}^{(j)}}{T}}\right]\right. \\ &  + \left.\left[\pi(a_{f,kj} = 2\mid \widehat{\mC}_k^{(j)})  2^{-2\frac{\mbE_{\pi_{l,k}}W_{kr}^{(j)}}{T}}\right]\right] + \lambda \Bigg. \Bigg) W_{kr}^{(j)}  = 0.
    \end{aligned}
    \label{eq_1}
    \vspace{-2mm}\eeq
Simplifying \eqref{eq_1}, $\pi(a_{l,kr} \mid \mC_k)$ can be written as:
\vspace{-2mm}\beq
\begin{aligned}
&\log_2\left(\left(\frac{\alpha_1}{J} + \lambda\right)\frac{T}{\alpha_2\sigma^2}\right) =  -2\sum\limits_{d}\pi(a_{l,kd} \mid \mC_k)\frac{W_{kd}^{(j)}}{T} \\ &+  \log2\left[ p(\widehat{c}_{kr}^{(j)}\mid \widehat{c}_{kr}^{(j)}) \left[1- \pi(a_{f,kj} = 3\mid \widehat{\mC}_k^{(j)}) \right] \right].
\end{aligned}
\eeq
Finally, we obtain $\pi(a_{l,kr} \mid \mC_k)$ as \eqref{eq_Wkj}.
For a given semantic relevance vector $\bmw_{k}^{(j)}$ and  $\pi(a_{l,kr} \mid \mC_k)$, $\lambda$ can be computed  using bisection such that:
\vspace{-2mm}\beq
\vspace{-5mm}\begin{aligned}
    &\sum\limits_{\forall a_{l,kr}}\pi(a_{l,kr}\mid \mC_k) = 1, \forall r, \,\, \,\textrm{and},\\\vspace{-2mm}
   &\mbE_{\pi_{l,k}} \sum\limits_{k\in \mK}\sum\limits_{c_{kr}\in\mC_{k}}\frac{1}{J}\sum\limits_{j \in \mJ}w_{kr}^{(j)} a_{l,kr}(\lambda) &\leq B.
\end{aligned}
\label{eq_bisection}
\vspace{-0mm}\eeq
\vspace{-0mm}\section{Proof of Theorem~\ref{Theorem1}}
\label{proof_theorem1}\vspace{-2mm}
    For a fixed perception, the TX strategy converges to the globally optimal solution of the utility function \eqref{eq_TX_utility}. For the RX utility function, the inverse QoTE considered is non-convex nature. The resulting linear approximation from Taylor series approximation as derived in Lemma~\ref{lemma3}. Hence, for the Stackelberg hypergame considered, the TX and RX strategies converge to a local Stackelberg equilibrium solution. As studied in Lemma~\ref{lemma_conv_mis}, given a fixed TX and RX strategies, the perceptions converge, with a monotonically decreasing misperception function. Hence, we can conclude that each of the user strategies $\pi(\bma_{s,i}\mid \mC), \forall i \in \{\mK,\mJ\}, s\in \{l,f\}$ converge to a local HSE.
       
\vspace{-0mm}\section{Proof of Lemma~\ref{lemma3}}
\label{proof_lemma3}\vspace{-1mm}

First, we fix $\pi(a_{f,kj}=0\mid \widehat{C}_k) = p(\abs{c_{kr}-\widehat{c}_{kr}}^2 \geq \delta)$, where $\delta$ is the maximum error that can be tolerated such that the semantic information conveyed by $\widehat{c}_{kr}$ is same as that of $c_{kr}$. We refer the readers to more details on how to define this radius $\delta$ to our previous work \cite{ChristoTWCArxiv2022}.
First, we look at deriving the expression for $\pi(a_{f,kj} = 1\mid \widehat{\mC}_k^{(j)})$. From Lemma~\ref{Lemma1}, it is clear that $P(\tau_{kj} \leq \tau^{\mathrm{max}})$ is non-convex function of $\beta_{kj}(1)$ and hence of $\pi(a_{f,kj} = 1\mid \widehat{\mC}_k^{(j)})$. Hence, we first perform a Taylor series approximation for $x_{kj}$ as follows.
     \vspace{-1mm}\beq
     x_{kj} \approx x_{kj}^0 + (\beta_{kj}(1) - \beta_{kj}(1)^{0})x_{kj}^{\prime}(\beta_{kj}(1)^0)
     \eeq
     Further, we obtain the derivative of the Lagrangian with respect to $\pi(a_{f,kj} = 1\mid \widehat{\mC}_k^{(j)})$ and equates to zero: 
     \beq
     \begin{aligned}
     &e^{-\frac{x_{kj}^2}{2}}\sum\limits_{r=1}^D p(\widehat{c}_{kr}^{(j)}\mid \widehat{c}_{kr}^{(j)}) \abs{c_{kr} - \widehat{c}_{kr}^{(j)}}^2 - \\   &x_{kj}^{\prime}(\beta_{kj}(1)^0)x_{kj}e^{-\frac{x_{kj}^2}{2}}\left(\sum\limits_{r=1}^D p(\widehat{c}_{kr}^{(j)}\mid \widehat{c}_{kr}^{(j)})\frac{F_{kr}^{(j)}}{F_{j}^{\mathrm{max}}}\right)\\ &\sum\limits_{r=1}^D p(\widehat{c}_{kr}^{(j)}\mid \widehat{c}_{kr}^{(j)})\pi(a_{f,kj} = 1\mid \widehat{\mC}_k^{(j)}) \abs{c_{kr} - \widehat{c}_{kr}^{(j)}}^2 +\gamma_{kj} F_{kr}^{(j)} = 0
     \end{aligned}
     \label{eq_der_RX_akj1}
     \eeq\vspace{-0mm}
From \eqref{eq_der_RX_akj1}, we compute $\pi(a_{f,kj} = s\mid \widehat{\mC}_k^{(j)}), d \in \{1,2\}$ via an iterative process as \eqref{eq_pi_akj_equi}.
      Finally, we obtain $\pi(a_{f,kj} = 3\mid \widehat{C}_k^{(j)})$ as \eqref{eq_pi_akj_3}.

     \vspace{-3mm}\section{Proof of Theorem~\ref{Theorem1}}
\label{proof_theorem1}\vspace{-2mm}
     For a fixed perception, the TX strategy converges to the globally optimal solution of the utility function \eqref{eq_TX_utility}. For the RX utility function, the inverse QoTE considered is non-convex nature. The resulting linear approximation from Taylor series approximation as derived in Lemma~\ref{lemma3}. Hence, for the Stackelberg hypergame considered, the TX and RX strategies converge to a local Stackelberg equilibrium solution. As studied in Lemma~\ref{lemma_conv_mis}, given a fixed TX and RX strategies, the perceptions converge, with a monotonically decreasing misperception function. Hence, we can conclude that each of the user strategies $\pi(\bma_{s,i}\mid \mC), \forall i \in \{\mK,\mJ\}, s\in \{l,f\}$ converge to a local HSE.
       

\vspace{-2mm}\bibliographystyle{IEEEbib}
\def\baselinestretch{1}
\bibliography{refs,semantics_ref}

\begin{thebibliography}{10}

\bibitem{ChaccourArxiv2022}
C.~Chaccour, W.~Saad, M.~Debbah, Z.~Han, and H.~V. Poor,
\newblock ``{Less data, more knowledge: buildingnext generation semantic communication networks},''
\newblock {\em IEEE Communications Surveys \& Tutorials}, Jun. 2024.

\bibitem{SaadProceedings2024}
W.~Saad, O.~Hashash, C.~K. Thomas, C.~Chaccour, M.~Debbah, N.~Mandayam, and Z.~Han,
\newblock ``{Artificial general intelligence (AGI)-native wireless systems: A journey beyond 6G},''
\newblock {\em arXiv preprint arXiv:2405.02336}, 2024.

\bibitem{LiuTCCN2023}
C.~Liu, C.~Guo, Y.~Yang, and N.~Jiang,
\newblock ``{Adaptable semantic compression and resource allocation for task-oriented communications },''
\newblock {\em IEEE Transactions on Cognitive Communications and Networking}, Dec. 2023.

\bibitem{YanGC2022}
L.~Yan, Z.~Qin, R.~Zhang, Y.~Li, and G.~Y. Li,
\newblock ``{QoE-aware resource allocation for semantic communication networks},''
\newblock in {\em Proceedings of the IEEE Global Communications Conference (GLOBECOM)}, Rio De Janeiro, Brazil, Dec. 2022.

\bibitem{ZhaoArxiv2024}
Z.~Zhao, Z.~Yang, M.~Chen, H.~V. Poor, and Z.~Zhang,
\newblock ``A joint communication and computation design for probabilistic semantic communications,''
\newblock {\em arXiv preprint arXiv:2402.16328}, 2024.

\bibitem{KovachGT2015}
N.~S. Kovach, A.~S. Gibson, and G.~B. Lamont,
\newblock ``{Hypergame theory: a model for conflict, misperception, and deception. Game Theory },''
\newblock {\em Game Theory}, 2015.

\bibitem{GharesifardCDC2010}
B.~Gharesifard and J.~Cort{\"e}s,
\newblock ``{Evolution of the perception about the opponent in hypergames},''
\newblock in {\em Proceedings of 49th IEEE Conference on Decision and Control (CDC)}, 2010, pp. 1076--1081.

\bibitem{ChaccourITJ2022}
C.~Chaccour, M.~N. Soorki, W.~Saad, M.~Bennis, and P.~Popovski,
\newblock ``{Can Terahertz Provide High-Rate Reliable Low-Latency Communications for Wireless VR},''
\newblock {\em IEEE Internet of Things Journal}, vol. 9, no. 12, Jun. 2022.

\bibitem{ChristoTWCArxiv2022}
C.~K. Thomas and W.~Saad,
\newblock ``{Neuro-Symbolic Causal Reasoning Meets Signaling Game for Emergent Semantic Communications},''
\newblock {\em IEEE Transactions on Wireless Communications}, vol. 23, no. 5, May. 2024.

\bibitem{PearlBasic2018}
J.~Pearl and D.~Mackenzie,
\newblock ``{The Book of Why},''
\newblock in {\em Basic Books}, 2018.

\bibitem{AgliettiPMLR2020}
V.~Aglietti, X.~Lu, A.~Paleyes, and J.~Gonz{\'a}lez,
\newblock ``{Causal Bayesian Optimization},''
\newblock in {\em Proceedings of the 23rd International Conference on Artificial Intelligence and Statistics (AISTATS)}, Jun. 2020.

\bibitem{ThomasGC2022}
C.~K. Thomas and W.~Saad,
\newblock ``{Neuro-Symbolic Artificial Intelligence (AI) for Intent based Semantic Communication},''
\newblock in {\em Proceedings of IEEE Global Communications Conference (GLOBECOM)}, Dec 2022.

\bibitem{CoverThomas1991}
T.~M. Cover and J.~A. Thomas,
\newblock ``{Elements of Information Theory},''
\newblock in {\em Wiley}, 1991.

\bibitem{ChengTIFS2022}
Z.~Cheng, G.~Chen, and Y.~Hong,
\newblock ``{Single-leader-multiple-followers Stackelberg security game with hypergame framework },''
\newblock {\em IEEE Transactions on Information Forensics and Security}, vol. 17, no. 2, 2022.

\end{thebibliography}

\end{document}